# Klein-Gordon equation particles in exponential-type molecule potentials and its thermodynamic properties in D- dimensions

*A.N. Ikot[1*], B.C.Lutfuoglu[2], M. I.Ngwueke[1], M.E.Udoh[1], S.Zare[3] and H. Hassanabadi[3]*

[1]Department of Physics, Theoretical Physics Group, University of Port Harcourt, Choba, P.M.B 5323 Port Harcourt-Nigeria
[2]Department of Physics, Akdentz University,07058 Antalya,Turkey
[3]Department of Physics, Shahrood University of technology, Shahrood, Iran.

*email:ndemikotphysics@gmail.com
Tel:+2348038767186*

**Abstract**

In this paper we use the Nikiforv-Uvarov method to obtain the approximate solutions of the Klein-Gordon equation with deformed five parameter exponential type potential (DFPEP) model. We also obtain the solutions of the Schrödinger equation in the presence of the DFPEP in the non-relativistic limits. In addition, we calculate in the non-relativistic limits the thermodynamics properties such as vibrational mean energy U, free energy F and the specific heat capacity C . Special cases of the potential are also discussed.

## 1. Introduction

The Nikiforov-Uvarov[1], exact quantization rule[2], asymptotic iteration method (AIM) [3] and Supersymmetric quantum mechanics (SUSYQM)[4] have been used by many authors to obtained the bound state solutions and scattering state solutions of Schrödinger, Klein-Gordon and Dirac equations with different potential models [5-10]. Also, the recent study of the relativistic wave equation in the recent years particularly the Klein-Gordon and Dirac equation have attracted the attention of many authors because of the solutions these equations play in getting the relativistic effect in nuclear physics and other areas [11]. The Klein-Gordon equation (KGE) is the well-known relativistic wave equation that describes spin-zero particles. It is also known that the analytic solutions of the Klein-Gordon equations are only possible in few cases such as harmonic and Coulomb potentials [12-14]. It is a common exercise in quantum mechanics that one usually looks for exact solutions which however, for arbitrary $l$ -states ($l \neq 0$), the Klein-

Gordon does not admit an exact solution [15-17]. Thus, the Klein-Gordon equation can only be solved approximately using different approximation schemes [18-20]. Improved Greene and Aldrich approximation has been used by Jia et al., [21] to investigate the bound state solution of the Schrödinger equation with Manning-Rosen potential. Ikot et al [22-23] studied the multiparameter exponential type potential with improved approximation in both relativistic and non-relativistic regimes. Recently, the Klein-Gordon, Dirac and Schrödinger equations in generalized D-dimensions for different potentials is getting the attention of many researchers [24-26]. These multidimensional space analyses of the Klein-Gordon, Dirac and Schrödinger equations have also been investigated for different potentials [27-30].

In this paper, we are interested in the deformed five parameter exponential type potential (DFPEP) proposed by Jia et al[31]. This potential is one of the most realistic potential models in nuclear physics, molecular physics, condensed matter physics and elementary particle physics since the potential incorporate the well-known potential models as special cases. The DFPEP model which incorporates these special well-known potentials can be used in the study of nuclear structure in the spherical shell model to obtain the energy splitting and the magic numbers [32]. Consequently, one can say authoritatively that the achievement of quantum mechanics in the description of the physics of atomic and sub-atomic particles can not be overemphasized [33-48].

The DFPEP is defined as [31]

$$V(r) = p_1 + \frac{p_2 e^{-2\alpha r}}{\left(1 - q e^{-2\alpha r}\right)} + \frac{p_3 e^{-4\alpha r}}{\left(1 - q e^{-2\alpha r}\right)^2} \quad (1)$$

where $p_i (i = 1...3)$ are the potential parameters and the range of q-parameter is $q > 0$ or $-1 \leq q < 0$.

In the present study, we intend to investigate the Klein-Gordon equation for the DFDEP potential within the framework of NU method in D-dimensions. This potential had been studied in the non-relativistic regime by Jia et al. [31] for s-wave and obtained the exact results using SUSYQM. However, because of the physics of the problem at hand, one can considered Klein-Gordon equation in D dimension. Thus, we will study the problem in special cases such as 1D, 2D, 3D, 4D, 5D and 6D in our computation in order to give a wider energy spectrum to see the behaviour of energy for large states. More recently many researchers in theoretical physics have studied different models in which the spacetime has more dimensions than the four dimensions observable in our daily experience. The some applications of these models are found in string theory [34]. Also the solutions of higher dimensional general relativity in physics are reported [35] and references therein.

In addition, the thermodynamics properties for some physical systems have been investigated [40-44]. Dong et al.[41] studied the hidden symmetries and thermodynamics properties for a harmonic oscillator plus inverse square potential. The thermodynamics properties of the Schrödinger equation with shifted Deng-Fan potential have been studied [40].

The purpose of this work is to use NU formalism to obtain the approximate energy levels and the corresponding wave function relativistic particles with DFDEP using the improved approximation scheme for the centrifugal term and study the thermal properties such as vibratonal partition function (Z),vibrational mean energy (U),free energy (F) and specific heat capacity (C) at the non-relativistic limits.

The paper is organized as follows. In section 2, we present the Klein-Gordon in D-dimensions and bound state solutions. Section 3 is devoted to the thermodynamics properties in the non-relativistic limits. In section 4, we present the energy eigenvalues of special cases. Section 5 is devoted to the discussion of the numerical results. Finally, we give a brief conclusion in section 6.

## 2. Solutions of Klein-Gordon Equation in D-Dimension

The Klein-Gordon equation in higher dimension for spherically symmetric potential reads [22-25],

$$-\Delta_D \psi_{n,l,m}(r,\Omega_D) = \left\{ [E_{n,l} - V(r)]^2 - [m + S(r)]^2 \right\} \psi_{n,l,m}(r,\Omega_D) \qquad (2)$$

Where $E_{n,l}, \mu, V(r)$ and $S(r)$ are the relativistic energy, rest mass, the repulsive vector potential and the attractive scalar potential respectively and $\Delta_D$ is defined as

$$\Delta_D = \nabla_D^2 = \frac{1}{r^{D-1}} \frac{\partial}{\partial} \left( r^{D-1} \frac{\partial}{\partial r} \right) - \frac{\Lambda_D^2(\Omega_D)}{r^2} \qquad (3)$$

The total wave function in D-dimension is written as,

$$\psi_{n,l,m}(r,\Omega_D) = R_{n,l}(r) Y_l^m(\Omega_D) \qquad (4)$$

The term $\frac{\Lambda_D^2(\Omega_D)}{r^2}$ is the generalization of the centrifugal term for the higher dimensional space. The eigenvalues of $\Lambda_D^2(\Omega_D)$ are defined by the relation,

$$\Lambda_D^2(\Omega_D) Y_l^m(\Omega_D) = l(l+D-2) Y_l^m(\Omega_D) \qquad (5)$$

Where $Y_l^m(\Omega_D), R_{n,l}$ and $l$ represent the hyperspherical harmonics, the hyperradial wave function and the orbital angular momentum quantum number respectively. Now substituting ansatz $R_{n,l}(r) = r^{-\frac{(D-1)}{2}} F_{n,l}(r)$ for the wave function into equation (4) yields,

$$\left\{ \frac{d^2}{dr^2} + \left(E_{n,l} - V(r)\right)^2 - \left(\mu + S(r)\right)^2 - \frac{(D+2l-1)(D+2l-3)}{4r^2} \right\} F_{n,l}(r) = 0 \qquad (6)$$

Now considering equal scalar and vector potentials $S(r) = V(r)$ in Eq. (6), we obtain the second order Schrodinger-like equation

$$\left\{ \frac{d^2}{dr^2} + E_{n,l}^2 - \mu^2 - 2(E_{n,l} + \mu)\left( p_1 + \frac{p_2 e^{-2\alpha r}}{(1-qe^{-2\alpha r})} + \frac{p_3 e^{-4\alpha r}}{(1-qe^{-2\alpha r})^2} \right) - \frac{(D+2l-1)(D+2l-3)}{4r^2} \right\} F_{n,l}(r) = 0 \qquad (7)$$

It is obvious that the Klein-Gordon equation above cannot be solved analytically for the DFPEP for $l \neq 0$. Thanks to Greene and Aldrich [18] who first proposed an approximation for the centrifugal terms and other researchers [36]. Following this development many authors have proposed an improved approximation scheme for the centrifugal term[37]. As a consequent, we adopt the proper approximation for the centrifugal term which is valid for both short and long range potential as [38],

$$\frac{1}{r^2} \approx \left( c_0 + \frac{\omega e^{-2\alpha r}}{(1-qe^{-2\alpha r})} + \frac{\lambda e^{-4\alpha r}}{(1-qe^{-2\alpha r})^2} \right) \qquad (8)$$

where $c_0 = \frac{1}{12}$ [38], $\omega$ and $\lambda$ are the two adjustable parameter. If $c_0 = 0, \omega = 0, q = 1$ and $\lambda \to 4\alpha^2$, then Eq.(8) turns to the approximation scheme suggested by Greene and Aldrich [18]. Now substituting Eq.(8) into equation (7) and with a new coordinate transformation $z = qe^{-2\alpha r}$, we obtain the following second order Schrödinger-like differential equation,

$$\frac{d^2 F_{nl}(z)}{dz^2} + \frac{1}{z}\frac{dF_{nl}(z)}{dz} + \frac{1}{z^2(1-z)^2}\left[ \frac{\gamma_1}{4\alpha^2}(1-z)^2 - \frac{\gamma_2}{4\alpha^2 q}z(1-z) - \frac{\gamma_3}{4\alpha^2 q^2}z^2 \right] F_{nl}(z) = 0 \qquad (9)$$

Where,

$$\gamma_1 = E^2 - \mu^2 - 2(E+\mu)p_1 - \frac{c_0(D+2l-3)(D+2l-1)}{4},$$

$$\gamma_2 = 2(E+\mu)p_2 + \frac{(D+2l-3)(D+2l-1)\omega}{4}, \quad (10)$$

$$\gamma_3 = 2(E+\mu)p_3 + \frac{(D+2l-3)(D+2l-1)\lambda}{4}$$

Or more explicitly, we can rewrite Eq.(9) in the form

$$\frac{d^2 F_{nl}(z)}{dz^2} + \frac{(1-z)}{z(1-z)}\frac{dF_{nl}(z)}{dz} + \frac{1}{z^2(1-z)^2}\{\omega_1 z^2 + \omega_2 z + \omega_3\} F_{nl}(z) = 0 \quad (11)$$

Where,

$$\omega_1 = \frac{\gamma_1}{4\alpha^2} + \frac{\gamma_2}{4\alpha^2 q} - \frac{\gamma_3}{4\alpha^2 q^2}$$

$$\omega_2 = -\left(\frac{2\gamma_1}{4\alpha^2} + \frac{\gamma_2}{4\alpha^2 q}\right) \quad (12)$$

$$\omega_3 = \frac{\gamma_1}{4\alpha^2}$$

Applying the NU method [1,39] (see appendix A.2), we obtain the following coefficients

$$\alpha_1 = \alpha_2 = \alpha_3 = 1, \xi_1 = -\omega_1, \xi_2 = \omega_2, \xi_3 = -\omega_3 \quad (13)$$

And using Eq.(A.5), we obtain the remaining coefficients as,

$$\alpha_4 = 0, \alpha_5 = -\frac{1}{2}, \alpha_6 = \frac{1}{4} - \frac{\gamma_1}{4\alpha^2} - \frac{\gamma_2}{4\alpha^2 q} + \frac{\gamma_3}{4\alpha^2 q^2}, \alpha_7 = \frac{2\gamma_1}{4\alpha^2} + \frac{\gamma_2}{4\alpha^2 q}, \alpha_8 = -\frac{\gamma_1}{4\alpha^2},$$

$$\alpha_9 = \frac{1}{4} + \frac{\gamma_3}{4\alpha^2 q^2}$$

$$\alpha_{10} = 1 + 2\sqrt{-\omega_3}, \alpha_{11} = 2 + 2\left(\sqrt{\frac{1}{4} + \frac{\gamma_3}{4\alpha^2 q^2}} + \sqrt{-\omega_3}\right),$$

$$\alpha_{12} = \sqrt{-\omega_3}, \alpha_{13} = -\frac{1}{2} - \left(\sqrt{\frac{1}{4} + \frac{\gamma_3}{4\alpha^2 q^2}} + \sqrt{-\omega_3}\right) \quad (14)$$

Thus, the energy eigenvalues can be obtain for the DFPEP using the NU method [1,39] (see appendix Eq.( A.4) ) as

$$E^2 - \mu^2 = -4\alpha^2 \left[ \frac{\left( n^2 + n + \frac{1}{2} + \frac{\gamma_2}{4\alpha^2 q} + (2n+1)\sqrt{\frac{1}{4} + \frac{\gamma_3}{4\alpha^2 q^2}} \right)^2}{\left( 2n + 1 + 2\sqrt{\frac{1}{4} + \frac{\gamma_3}{4\alpha^2 q^2}} \right)^2} \right] + 2(E+\mu)P_1 \quad (15)$$

$$+ \frac{(D+2l-1)(D+2l-3)c_0}{4}$$

The corresponding wave function for the DFPEP of the Klein-Gordon particles is obtain as follows,

$$F_{nl}(r) = \left( qe^{-2\alpha r} \right)^{\sqrt{-\frac{\gamma_1}{4\alpha^2}}} \left( 1 - qe^{-2\alpha r} \right)^{\frac{1}{2}\left( 1 + \sqrt{\frac{1}{4} + \frac{\gamma_3}{4\alpha^2 q^2}} \right)} P_n^{\left( 2\sqrt{-\frac{\gamma_1}{4\alpha^2}}, 2\sqrt{\frac{1}{4} + \frac{\gamma_3}{4\alpha^2 q^2}} \right)} \left( 1 - 2qe^{-2\alpha r} \right) \quad (16)$$

and

$$R_{nl}(r) = r^{-\left( \frac{D-1}{2} \right)} F_{nl}(r) \quad (17)$$

## 3 Thermodynamic properties of the non-relativistic Schrödinger equation with DFPEP

If we map $E + \mu \to \frac{2m}{\hbar^2}, E - \mu \to E_{nl}$, we obtain solution in the non-relativistic limit of the DFPEP from equation (15) as,

$$E_{nl} = -\frac{2\hbar^2 \alpha^2}{m} \left[ \frac{\left( n^2 + n + \frac{1}{2} + \frac{\gamma_2}{4\alpha^2 q} + (2n+1)\sqrt{\frac{1}{4} + \frac{\gamma_3}{4\alpha^2 q^2}} \right)^2}{\left( 2n + 1 + 2\sqrt{\frac{1}{4} + \frac{\gamma_3}{4\alpha^2 q^2}} \right)^2} \right] + 2P_1 \quad (18)$$

$$+ \frac{\hbar^2 (D+2l-1)(D+2l-3)c_0}{8m}$$

In order to evaluate the thermodynamics properties of the DFPEP model, we recast Eq.(16) in the form,

$$E_{nl} = \frac{2\hbar^2 \alpha^2}{m} \left[ Q_1 - \left( \frac{Q_2}{2(n+\sigma)} + \frac{n+\sigma}{2} \right)^2 \right] \quad (19)$$

Where,

$$Q_1 = \frac{2mp_1}{\hbar^2\alpha^2} + \frac{(D+2l-1)(D+2l-3)c_0}{16\hbar\alpha^2},$$

$$Q_2 = -\left(\frac{\gamma_2}{4\alpha^2 q} + \frac{\gamma_3}{4\alpha^2 q^2}\right), \sigma = \frac{1}{2}\left(1 + \sqrt{1 + \frac{\gamma_3}{\alpha^2 q^2}}\right) \quad (20)$$

and $n = 0, 1, 2... < \left[-\sigma + \sqrt{Q_1} \pm \sqrt{Q_1 - Q_2}\right]$. However for the s-wave that is, $l = 0$ and $D = 0$, we obtain

$$E_{nl} = -\frac{2\hbar^2\alpha^2}{m}\left[\frac{\left(n^2 + n + \frac{1}{2} + \frac{\tilde{\gamma}_2}{4\alpha^2 q} + (2n+1)\sqrt{\frac{1}{4} + \frac{\tilde{\gamma}_3}{4\alpha^2 q^2}}\right)^2}{\left(2n + 1 + 2\sqrt{\frac{1}{4} + \frac{\tilde{\gamma}_3}{4\alpha^2 q^2}}\right)^2}\right] + 2P_1 \quad (21)$$

$$+ \frac{3\hbar^2 c_0}{8m}$$

$$\tilde{\gamma}_1 = \frac{2mE_{nl}}{\hbar^2} - \frac{4mp_1}{\hbar^2} - \frac{3c_0}{4},$$

$$\tilde{\gamma}_2 = \frac{4mp_2}{\hbar^2} + \frac{3\omega}{4}, \quad (22)$$

$$\tilde{\gamma}_3 = \frac{4mp_3}{\hbar^2} + \frac{3\lambda}{4}$$

### 3.1 Partition function

The vibrational partition function for the DFDEP model is given as [40-44],

$$Z_{vib}(\beta) = \sum_{n=0}^{\lambda} e^{-\beta E_{nl}}, \beta = \frac{1}{kT} \quad (23)$$

where $\eta = -\sigma + \sqrt{Q_1} \pm \sqrt{Q_1 - Q_2}$, $k$ is the Boltzmann constant and $T$ is the absolute temperature. Substituting Eq.(20) into Eq.(23), we obtain the partition function for the potential model as,

$$Z_{vib}(\beta) = \sum_{n=0}^{\eta} e^{A\beta + \frac{\beta B}{(n+\sigma)^2} + \beta C(n+\sigma)^2} \quad (24)$$

where,

$$A = \frac{2\hbar^2\alpha^2}{m}\left(\frac{Q_2}{2} - Q_1\right), B = \frac{\hbar^2\alpha^2 Q_2^2}{2m}, C = \frac{\hbar^2\alpha^2}{2m} \qquad (25)$$

In the classical limit, we can replaced the sum by an integral as,

$$Z_{vib}(\beta) = \int_0^\eta e^{A\beta + \frac{\beta B}{\rho^2} + \beta C \rho^2} d\rho, \rho = n + \sigma$$

$$= \frac{1}{2} e^{\beta C \rho^2 + \beta A} \sqrt{\beta B} \left[ \frac{2\eta e^{\frac{B}{\eta^2}}}{\sqrt{\beta B}} - \frac{2\sqrt{\beta B}\sqrt{\pi} \, erfi\left(\frac{\sqrt{\beta B}}{\eta}\right)}{\sqrt{\beta B}} - 2\sqrt{\pi} \right] \qquad (26)$$

**3.2 Vibrational mean energy U**

The vibrational mean free energy for the DFPEP model is obtain as,

$$U(\beta,\lambda) = -\frac{\partial}{\partial \beta} \ell n Z_{vib}(\beta,\lambda)$$

$$= -\left(\frac{2\delta_1 - \frac{1}{4}\delta_2 + \delta_3}{\delta_4}\right) \qquad (27)$$

where,

$$\delta_1 = \frac{1}{2} e^{\beta C \rho^2 + \beta A} \left( C\rho^2 + A \right) \sqrt{\beta B} \left( \frac{2\eta e^{\frac{B}{\eta^2}}}{\sqrt{\beta B}} - 2\sqrt{\beta B}\sqrt{\pi} \, erfi\left( \frac{\sqrt{\beta B}}{\eta} \right) - 2\pi \right),$$

$$\delta_2 = \frac{e^{\beta C\rho^2 + \beta A} \left( \frac{2\eta e^{\frac{B}{\eta^2}}}{\sqrt{\beta B}} - \frac{2\sqrt{\beta B}\sqrt{\pi}\, erfi\left( \frac{\sqrt{\beta B}}{\eta} \right)}{\sqrt{\beta B}} - 2\sqrt{\pi} \right) B}{\sqrt{\beta B}}$$

$$\delta_3 = \frac{1}{2} e^{\beta C\rho^2 + \beta A} \sqrt{\beta B} \left( \frac{\eta e^{\frac{B}{\eta^2}}}{(\beta B)^{\frac{3}{2}}} - \frac{\sqrt{\pi}\, erfi\left( \frac{\sqrt{\beta B}}{\eta} \right) B}{\sqrt{\beta B}\sqrt{\beta B}} - \frac{2Be^{\frac{B}{\eta^2}}}{\eta\sqrt{\beta B}} - \frac{\sqrt{\beta B}\sqrt{\pi}\, erfi\left( \frac{\sqrt{\beta B}}{\eta} \right) B}{(\beta B)^{\frac{3}{2}}} \right),$$

$$\delta_4 = e^{\beta C\rho^2 + \beta A} \sqrt{\beta B} \left( \frac{2\eta e^{\frac{B}{\eta^2}}}{(\beta B)^{\frac{1}{2}}} - \frac{2\sqrt{\beta B}\sqrt{\pi}\, erfi\left( \frac{\sqrt{\beta B}}{\eta} \right)}{\sqrt{\beta B}} - 2\sqrt{\pi} \right) \quad (28)$$

### 3.3 Vibrational mean free energy F

$$F(\beta) = -kT \ln Z_{vib}(\beta)$$

$$= -\frac{1}{\beta} \ln \left( \frac{1}{2} e^{\beta C\rho^2 + \beta A} \sqrt{\beta B} \left[ \frac{2\eta e^{\frac{B}{\eta^2}}}{\sqrt{\beta B}} - \frac{2\sqrt{\beta B}\sqrt{\pi}\, erfi\left( \frac{\sqrt{\beta B}}{\eta} \right)}{\sqrt{\beta B}} - 2\sqrt{\pi} \right] \right) \quad (29)$$

### 3.4 Vibrational specific heat C

$$C(\beta) = -\frac{\partial}{\partial \beta} \ln Z_{vib}(\beta)$$

$$= k\beta^2 \frac{\partial}{\partial \beta} \left( \frac{2\Lambda_1 + \Lambda_2 + \Lambda_3}{\Lambda_4} \right) \quad (30)$$

where,

$$\Lambda_1 = \frac{1}{2}\left( C\rho^2 + A \right) e^{\beta C\rho^2 + \beta A} \sqrt{\beta B} \left( \frac{2\eta e^{\frac{\beta B}{\eta^2}}}{\sqrt{\beta B}} - \frac{2\sqrt{B}\sqrt{\pi}\, erfi\left( \frac{\sqrt{\beta B}}{\eta} \right)}{\sqrt{\beta B}} - 2\sqrt{\pi} \right) \quad (31)$$

$$\Lambda_2 = \frac{1}{4} e^{\beta C \rho^2 + \beta A} \left( \frac{2\eta e^{\frac{\beta B}{\eta^2}}}{\sqrt{\beta B}} - \frac{2\sqrt{B}\sqrt{\pi}\, erfi\left(\frac{\sqrt{\beta B}}{\eta}\right)}{\sqrt{\beta B}} - 2\sqrt{\pi} \right) \frac{B}{\sqrt{\beta}} \qquad (32)$$

$$\Lambda_3 = \frac{1}{2} e^{\beta C \rho^2 + \beta A} \sqrt{\beta B} \left( \frac{2Be^{\frac{\beta B}{\eta^2}}}{\eta \sqrt{\beta B}} - \frac{\eta B e^{\frac{\beta B}{\eta^2}}}{(\beta B)^{\frac{3}{2}}} - \frac{2\sqrt{B} e^{\frac{\beta B}{\eta^2}}}{\eta \beta} + \frac{B^{\frac{3}{2}} \sqrt{\pi}\, erfi\left(\frac{\sqrt{\beta B}}{\eta}\right)}{(\beta B)^{\frac{3}{2}}} \right) \qquad (33)$$

$$\Lambda_4 = e^{\beta C \rho^2 + \beta A} \sqrt{\beta B} \left( \frac{2\eta e^{\frac{\beta B}{\eta^2}}}{\sqrt{\beta B}} - \frac{2\sqrt{B}\sqrt{\pi}\, erfi\left(\frac{\sqrt{\beta B}}{\eta}\right)}{\sqrt{\beta B}} - 2\sqrt{\pi} \right) \qquad (34)$$

**4 Few special cases of DFDEP**

In section, we will examine the special cases of DFDEP and compare our results with those found in the available literature.

**4.1 Manning-Rosen potential**

If we set $p_1 = 0, p_2 = -\frac{V_0}{b^2}, p_3 = \frac{\beta(\beta-1)}{b^2}, \alpha = \frac{1}{2b}, q = 1$ then the DFDEP turns into Manning-Rosen potential as [45],

$$V(r) = -\frac{V_0}{b^2} \frac{e^{-\frac{r}{b}}}{\left(1 - e^{-\frac{r}{b}}\right)} + \frac{\beta(\beta-1) e^{-\frac{2r}{b}}}{\left(1 - e^{-\frac{r}{b}}\right)^2} \qquad (35)$$

By substituting the corresponding parameters into Eqs.(15) and (18), we obtain the approximate energy spectra for the Manning-Rosen potential in the relativistic as

$$E^2 - \mu^2 = -\alpha^2 \left[ \frac{\Lambda^{MR}}{(n + \sigma^{MR})} + (n + \sigma^{MR}) \right]^2 + \frac{c_0 (D + 2l - 3)(D + 2l - 1)}{4}, \qquad (36)$$

where,

$$\Lambda^{MR} = b^2 \left( -\frac{2(E + \mu)}{b^2} (V_0 + \beta(\beta - 1)) + \frac{(D + 2l - 3)(D + 2l - 1)(\omega - \lambda)}{4} \right),$$

$$\sigma^{MR} = \frac{1}{2} \left( 1 + \sqrt{1 + 4b^2 \left( \frac{2\beta(\beta-1)}{b^2} + \frac{(D+2l-3)(D+2l-1)\lambda}{4} \right)} \right) \qquad (37)$$

and that in the non-relativistic limits as,

$$E_{nl} = -\frac{\hbar^2}{4mb^2}\left[\frac{\tilde{\Lambda}^{MR}}{(n+\tilde{\sigma}^{MR})}+(n+\tilde{\sigma}^{MR})\right]^2 + \frac{\hbar^2 c_0 (D+2l-1)(D+2l-3)}{4m} \quad (38)$$

where

$$\tilde{\Lambda}^{MR} = b^2\left(-\frac{4m}{\hbar^2 b^2}(V_0+\beta(\beta-1))+\frac{(D+2l-3)(D+2l-1)(\omega-\lambda)}{4}\right),$$

$$\tilde{\sigma}^{MR} = \frac{1}{2}\left(1+\sqrt{1+4b^2\left(\frac{4m\beta(\beta-1)}{\hbar^2 b^2}+\frac{(D+2l-3)(D+2l-1)\lambda}{4}\right)}\right) \quad (39)$$

These results are consistent with those obtained in refs.[45-46]

**4.2 Hulthen Potential**

For $p_1 = p_3 = 0, p_2 = -V_0\delta, q=1, \alpha = \frac{\delta}{2}$ then the DFDEP reduces to the Hulthen potential[48],

$$V(r) = -V_0\delta \frac{e^{-\delta r}}{1-e^{-\delta r}} \quad (40)$$

Now substituting these parameters into Eq.(15), we obtain the energy levels for the relativistic Hulthen potential as,

$$E^2 - \mu^2 = -\frac{\delta^2}{4}\left[\frac{\Lambda^H}{(n+\sigma^H)}+(n+\sigma^H)\right]^2 + \frac{c_0(D+2l-3)(D+2l-1)}{4}, \quad (41)$$

Where,

$$\Lambda^H = \frac{1}{\delta^2}\left(-2V_0(E+\mu)\delta + \frac{(D+2l-3)(D+2l-1)(\omega-\lambda)}{4}\right),$$

$$\sigma^H = \frac{1}{2}\left(1+\sqrt{1+\frac{4}{\delta^2}\left(\frac{(D+2l-3)(D+2l-1)\lambda}{4}\right)}\right) \quad (42)$$

Under the conditions, $\delta \to 0, \omega = \lambda = 1, c_0 = 0$, the energy spectrum of the Hulthen potential reduces to that of the Coulomb potential in the relativistic limit as

$$E_{nl} = -\mu\frac{(n+\sigma^H)^2 + V_0^2}{(n+\sigma^H)^2 - V_0^2} \quad (43)$$

The non-relativistic limit for the Hulthen potential is obtain from Eq.(18) as

$$E_{nl} = -\frac{\hbar^2\delta^2}{4m}\left[\frac{\tilde{\Lambda}^H}{(n+\tilde{\sigma}^H)}+(n+\tilde{\sigma}^H)\right]^2 + \frac{\hbar^2 c_0(D+2l-1)(D+2l-3)}{4m} \quad (44)$$

Where

$$\tilde{\Lambda} = \frac{1}{\delta^2}\left(-\frac{4mV_0\delta}{\hbar^2} + \frac{(D+2l-3)(D+2l-1)(\omega-\lambda)}{4}\right),$$

$$\tilde{\sigma} = \frac{1}{2}\left(1 + \sqrt{1 + \frac{4}{\delta^2}\left(\frac{(D+2l-3)(D+2l-1)\lambda}{4}\right)}\right) \quad (45)$$

However, when $\delta \to 0, \omega = \lambda = 1, c_0 = 0$, the energy spectrum of the Hulthen potential reduces to that of the Coulomb potential in the non-relativistic limit as,

$$E_{nl} = -\frac{4m\hbar^2 V_0^2}{\left(n + \sigma^H\right)^2} \quad (46)$$

These results are equivalent to those found in refs.[47-48].

## 5 Numerical Results

We have calculated the vibrational bound state energies of DFDEP potential for various $n$ and $l$ quantum numbers in $D$-dimensions. The calculated energy values are presented in Table 1 for $n = 0-2, l = 0-4$ and $D = 1-6$ in order to give a wider energy spectrum and its behaviour for various states in different dimensions. In fig.1, we plot the behaviour of the relativistic vibrational energy with the potential parameter $p_2$ for $D = 3, q = 1, p_1 = p_2 = 2$. The variations of the relativistic energy with other potential parameters are shown in fig.2 and fig.3 respectively. The plots of the wave function as a function of $r$ for various values of the potential parameters are shown in fig.4, fig.5 and fig.6 respectively.

We have also investigated the thermodynamics properties of DFDEP potential. This is achieved by first calculating the high temperature partition function and then obtained thermodynamics properties vis-à-vis internal mean vibrational energy $U(\beta)$, free energy $F(\beta)$ and specific heat $C_v(\beta)$.

For a given values of $\eta = 2, 4$ and 6 the dependence of the partition function $Z$ on $\beta$ is shown in fig.7 and this shows the effect of $\eta$ on the vibrational partition. The partition function decreases as $\eta$ is increase for high temperature $\beta$. Fig.8 shows the variation of the partition function with D-dimension for a low temperature regime $\beta = 0.005$ with $\eta = 2, 4$ and 6. This shows that the partition increases with increase in potential parameters $\eta$ and D. The variation of the vibrational partition functions as function of D at high temperature $\beta = 50$ at different $\eta(2, 4 \text{ and } 6)$ is shown in fig.9. At higher temperature, the partition tends to zero for higher values of $\eta = 4$ and 6 and the partition is dominant at $\eta = 2$.

We also show in fig.10 and fig.11, the behaviour of the vibrational partition function as function of the potential parameter $p_2$ at lower and higher temperature regimes. Fig.12 shows that the mean internal energy decreases monotonically with increasing $\beta$ as potential parameter is increase from $\eta = 20, 30$ and 50. In fig.13, we plot the free energy F as function of $\beta$ for different potential parameter $\eta = 20, 30$ and 50. Fig.13 shows that the vibrational free energy increases monotonically with increasing parameters $\eta$ and $\beta$.

Fig.14 shows that the specific heat capacity decreases monotonically with decreasing $\eta$ but with increasing $\beta$.

## 6 Conclusions

In this paper, we have solved the Klein-Gordon equation with DFDEP potential within the framework of the NU method. We obtained in details the energy eigenvalues and the corresponding wave function in D-dimensions. Special cases of the potential is deduced which are consistent with those found in the available literature. We also investigated the non-relativistic Schrödinger equation with DFDEP as special case of the relativistic Klein-Gordon equation. Under the non-relativistic regime, we study the thermodynamics properties of the DFDEP potential model. On the other hand, we have calculated the thermodynamics properties such as partition function ,vibrational mean energy,free energy and specific heat capacity.We also shown the behaviour of the thermodynamics properties as a function of the potential parameters.

**Appendix A**

**Basic concept of Nikiforov-Uvarov method**

The Nikiforov-Uvarovmethod[1,39] was proposed to solve second-order differential equation of the form

$$\psi''(s) + \frac{\tilde{\tau}(s)}{\sigma(s)}\psi'(s) + \frac{\tilde{\sigma}(s)}{\sigma^2(s)}\psi(s) = 0, \qquad (A.1)$$

With appropriate co-ordinate transformation, $s = s(r)$, where $\sigma(s)$ and $\tilde{\sigma}(s)$ are polynomial at most a second order and $\tilde{\tau}(s)$ is a first degree polynomial. The parametric form of the Schrodinger-like equation can be written for any potential as [39],

$$\frac{d^2\psi}{ds^2} + \frac{\alpha_1 - \alpha_2 s}{s(1-\alpha_3 s)}\frac{d\psi}{ds} + \frac{1}{s^2(1-\alpha_3 s)^2}\left[-\xi_1 s^2 + \xi_2 s - \xi_3\right]\psi(s) = 0 \qquad (A.2)$$

According to NU method, the eigenfunction and the corresponding energy eigenvalues equation becomes,

$$\psi(s) = s^{\alpha_{12}}(1-\alpha_3 s)^{-\alpha_{12}-\frac{\alpha_{13}}{\alpha_3}} P_n^{\left(\alpha_{10}-1,\frac{\alpha_{11}}{\alpha_3}-\alpha_{10}-1\right)}(1-\alpha_3 s) \qquad (A.3)$$

$$(\alpha_2 - \alpha_3)n + \alpha_3 n^2 - (2n+1)\alpha_5 + (2n+1)\left(\sqrt{\alpha_9} + \alpha_3\sqrt{\alpha_8}\right) + \alpha_7 + 2\alpha_3\alpha_8 + 2\sqrt{\alpha_8\alpha_9} = 0 \qquad (A.4)$$

Where

$$\alpha_4 = \frac{1-\alpha_1}{2}, \alpha_5 = \frac{(\alpha_2-2\alpha_3)}{2}, \alpha_6 = \alpha_5^2 + \xi_1, \alpha_7 = 2\alpha_4\alpha_5 - \xi_2,$$

$$\alpha_8 = \alpha_4^2 + \xi_3, \alpha_9 = \alpha_3\alpha_7 + \alpha_3^2\alpha_8 + \alpha_6, \alpha_{10} = \alpha_1 + 2\alpha_4 + 2\sqrt{\alpha_8},$$

$$\alpha_{11} = \alpha_2 - 2\alpha_5 + 2\left(\sqrt{\alpha_9} + \alpha_3\sqrt{\alpha_8}\right), \alpha_{12} = \alpha_4 + \sqrt{\alpha_8}, \alpha_{13} = \alpha_5 - \left(\sqrt{\alpha_9} + \alpha_3\sqrt{\alpha_8}\right) \qquad (A.5)$$

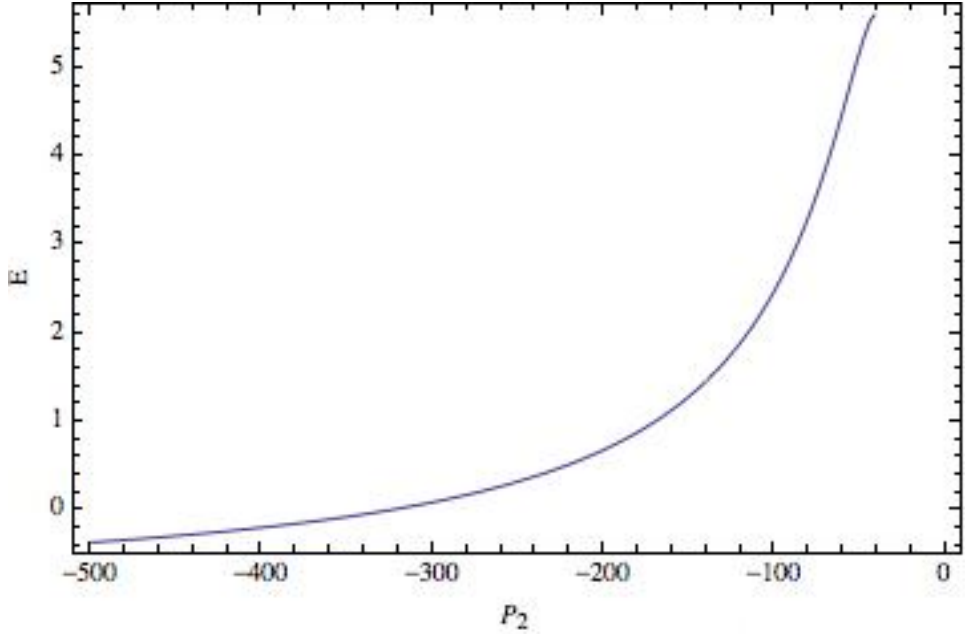

Figure 1. Behavior of $E$ versus $P_2$ varying for $C_0 = 2, \omega = -3, \alpha = -3, \lambda = 4\alpha^2,$
$D = 3, \mu = 1, q = 1, n = 2, l = 1, P_1 = 2, P_3 = 2$

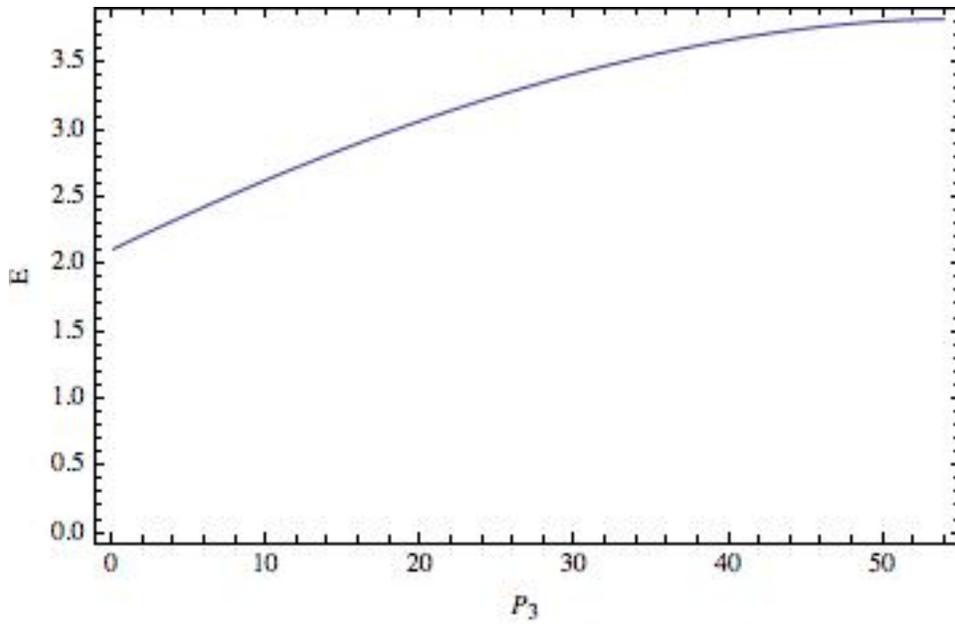

Figure 2. Behavior of $E$ versus $P_3$ varying for $C_0 = 2, \omega = -3, \alpha = -3, \lambda = 4\alpha^2,$
$D = 3, \mu = 1, q = 1, n = 2, l = 1, P_1 = 1, P_2 = -100$

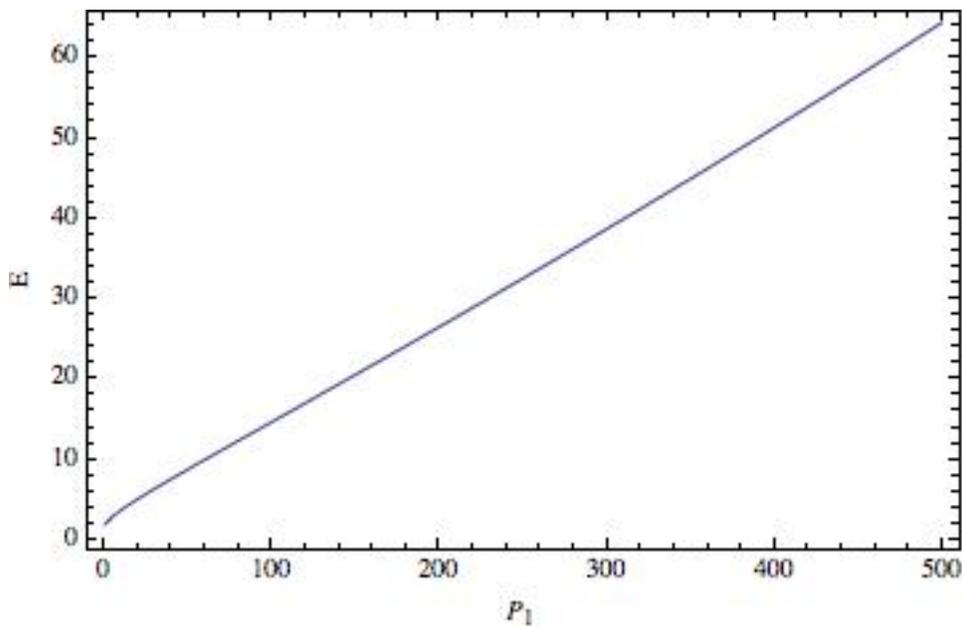

Figure 3. Behavior of $E$ versus $P_1$ varying for $C_0 = 2, \omega = -3, \alpha = -3, \lambda = 4\alpha^2,$
$D = 3, \mu = 1, q = 1, n = 2, l = 1, P_2 = -100, P_3 = 0.2$

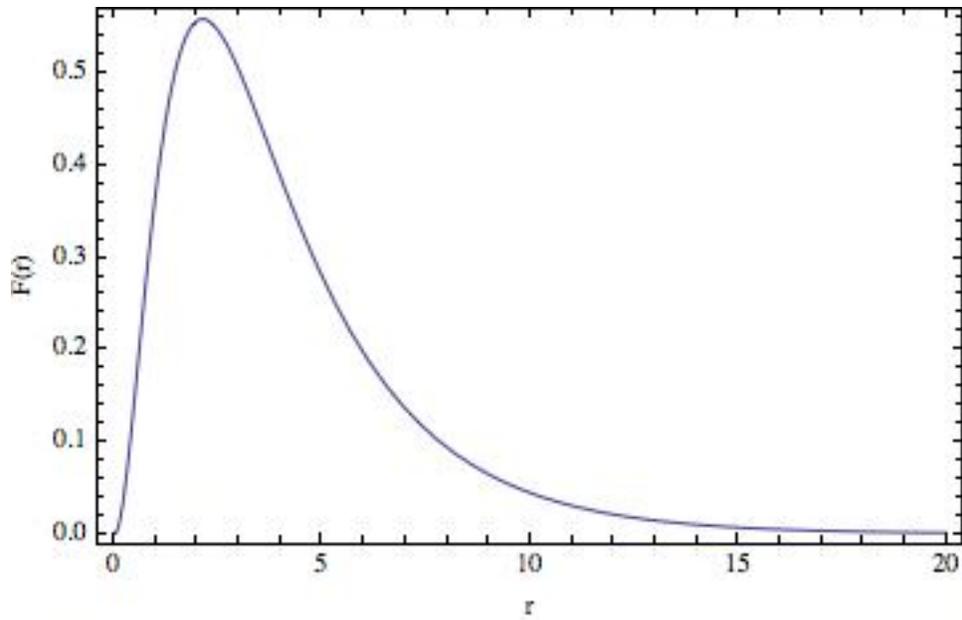

Figure 4. Behavior of $F(r)$ versus $r$ varying for $C_0 = 1, \omega = 1, \alpha = 0.4, \lambda = 0.01,$ $D = 3, \mu = 1, q = 1, n = 0, l = 1, P_1 = 2, P_2 = 1, P_3 = 0.5, \hbar = 1, E = 5.29524$

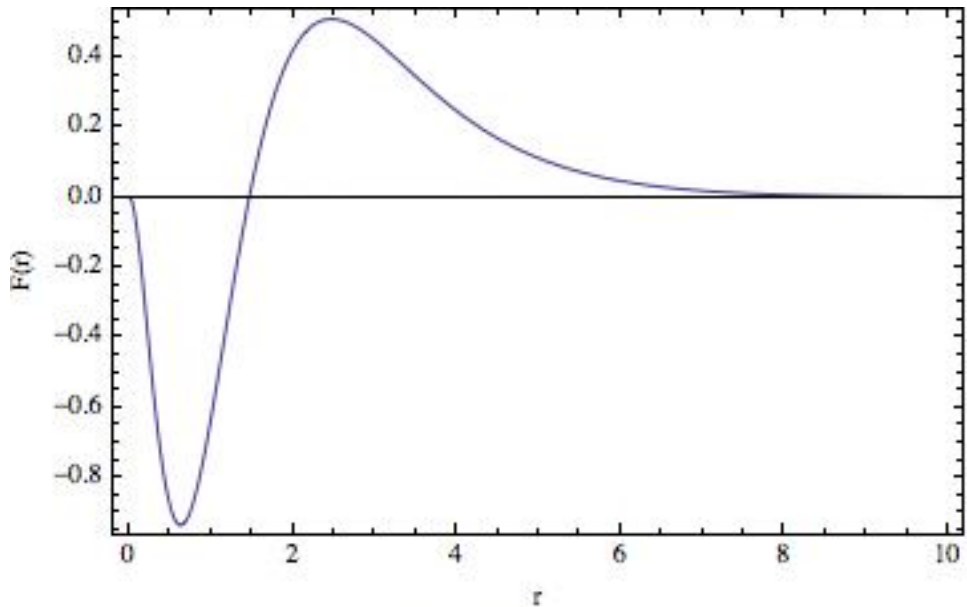

Figure 5. Behavior of $F(r)$ versus $r$ varying for $C_0 = 1, \omega = 1, \alpha = 0.4, \lambda = 0.01,$ $D = 3, \mu = 1, q = 1, n = 1, l = 1, P_1 = 2, P_2 = 1, P_3 = 0.5, \hbar = 1, E = 5.18775$

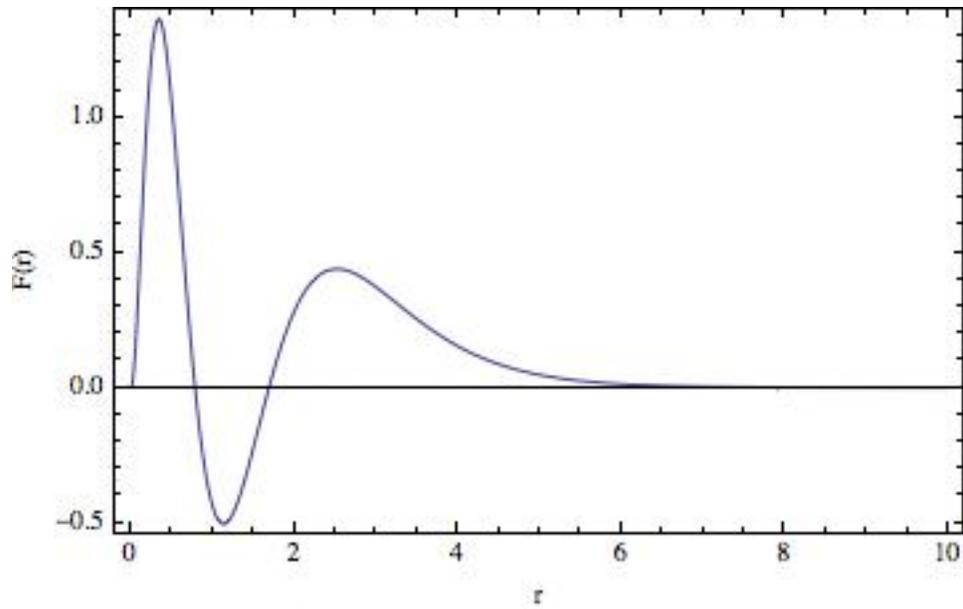

Figure 6. Behavior of $F(r)$ versus $r$ varying for $C_0 = 1, \omega = 1, \alpha = 0.4, \lambda = 0.01,$ $D = 3, \mu = 1, q = 1, n = 2, l = 1, P_1 = 2, P_2 = 1, P_3 = 0.5, \hbar = 1, E = 5.02272$

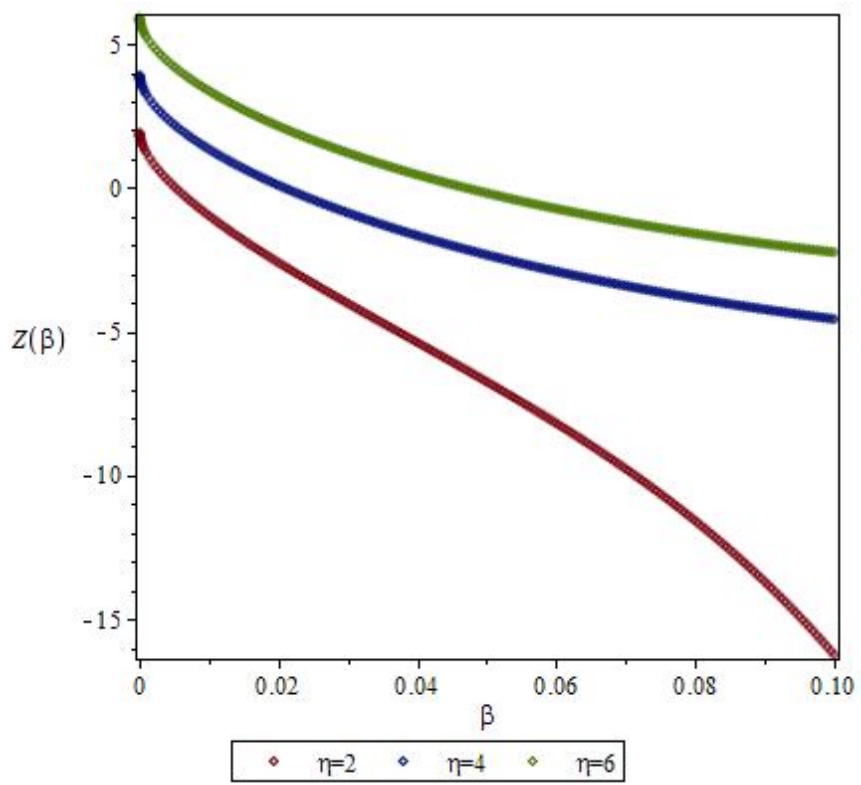

Fig.7: Vibrational partition function Z as a function of $\beta$ for different $\eta\,(2, 4\text{ and }6)$.

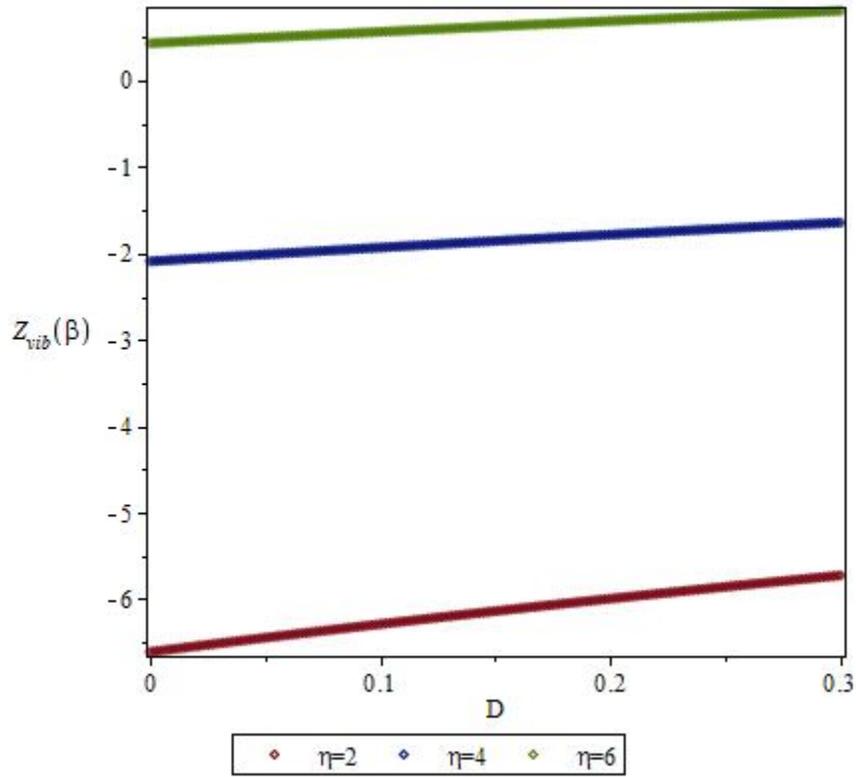

Fig.8: Variation of the partition function as a function of D-dimensions for different $\eta\,(2, 4\text{ and }6)$ at $\beta = 0.005$

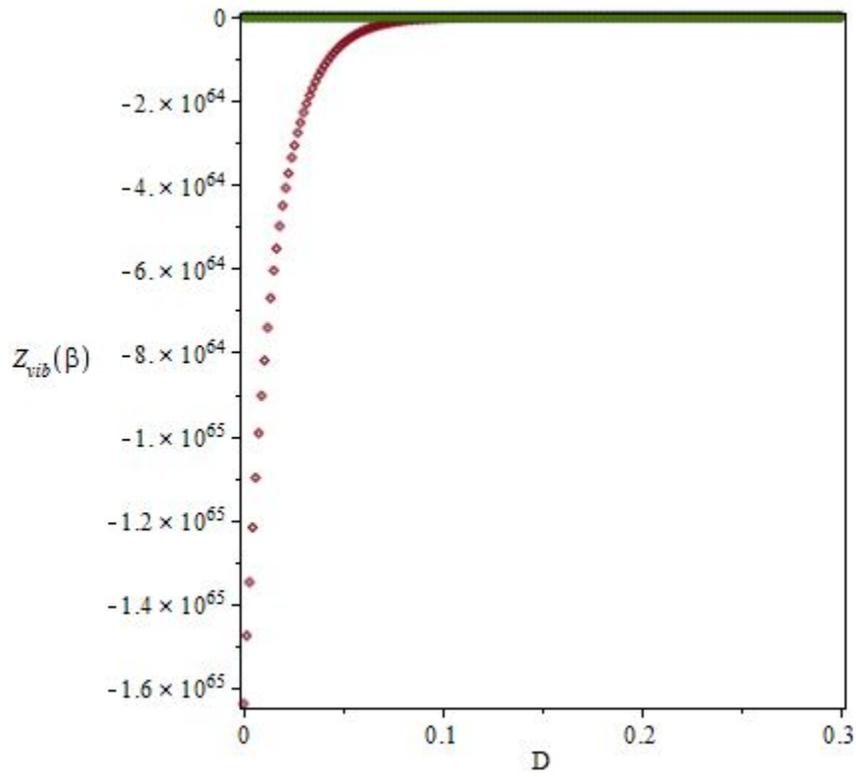

Fig.9: Partition function as function of D for different $\eta(2, 4 \text{ and } 6)$ at $\beta = 50$.

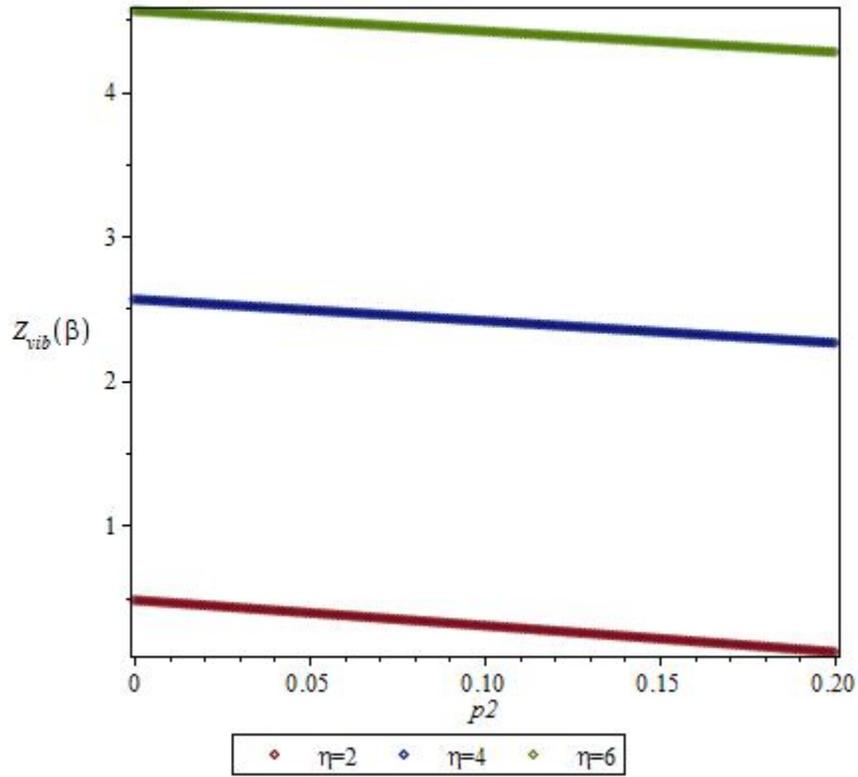

Fig.10: Variation of the partition function as a function of $p_2$ for different $\eta\,(2, 4 \text{ and } 6)$ at $\beta = 0.005$

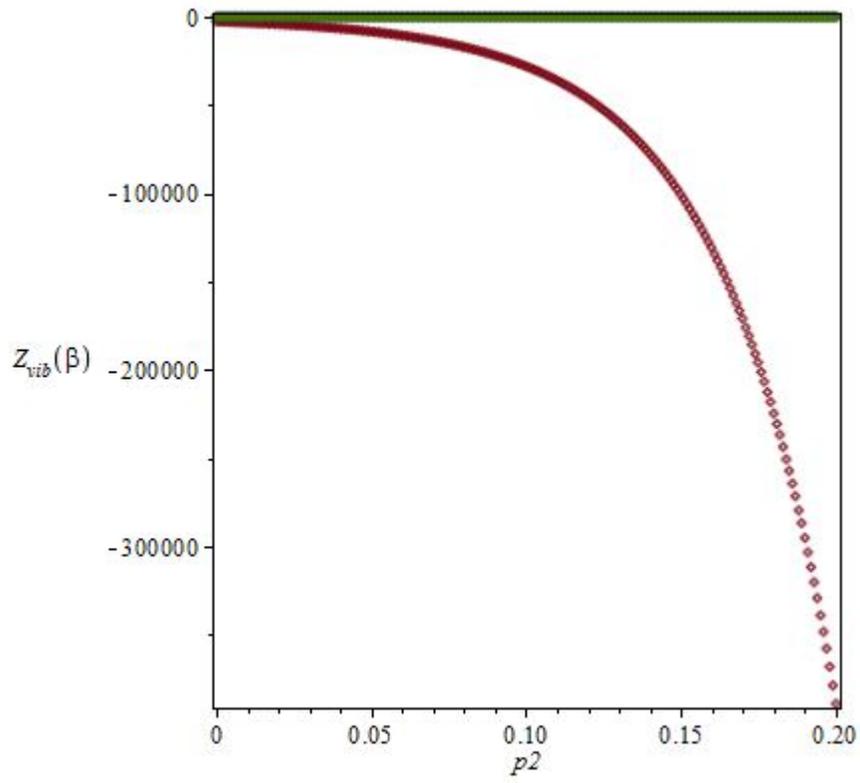

Fig.11: Partition function as function of $p_2$ for different $\eta\,(2, 4 \text{ and } 6)$ at $\beta = 50$.

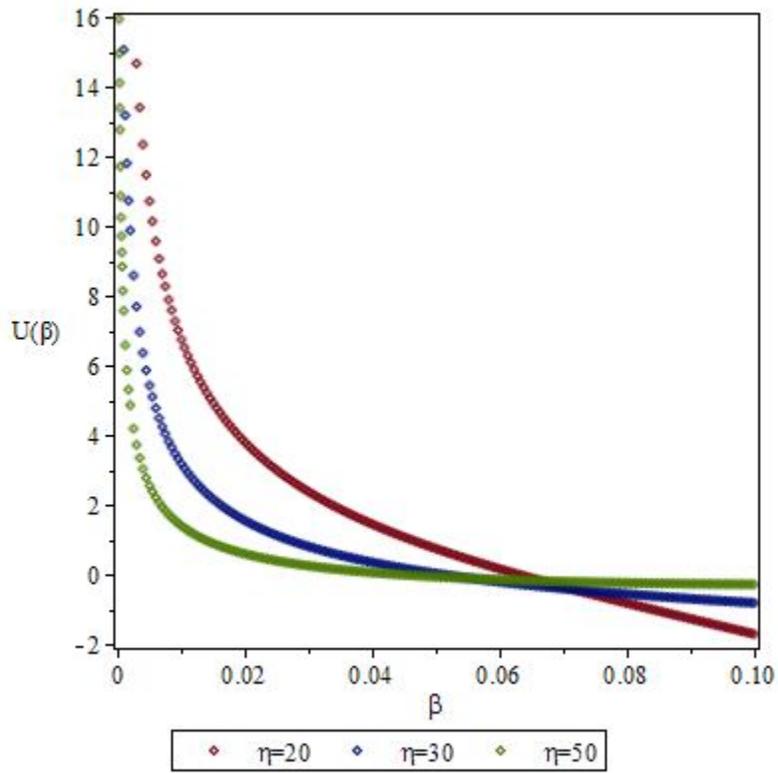

Fig.12: Vibrational mean energy U as function of $\beta$ for different $\eta\,(20, 30 \text{ and } 50)$.

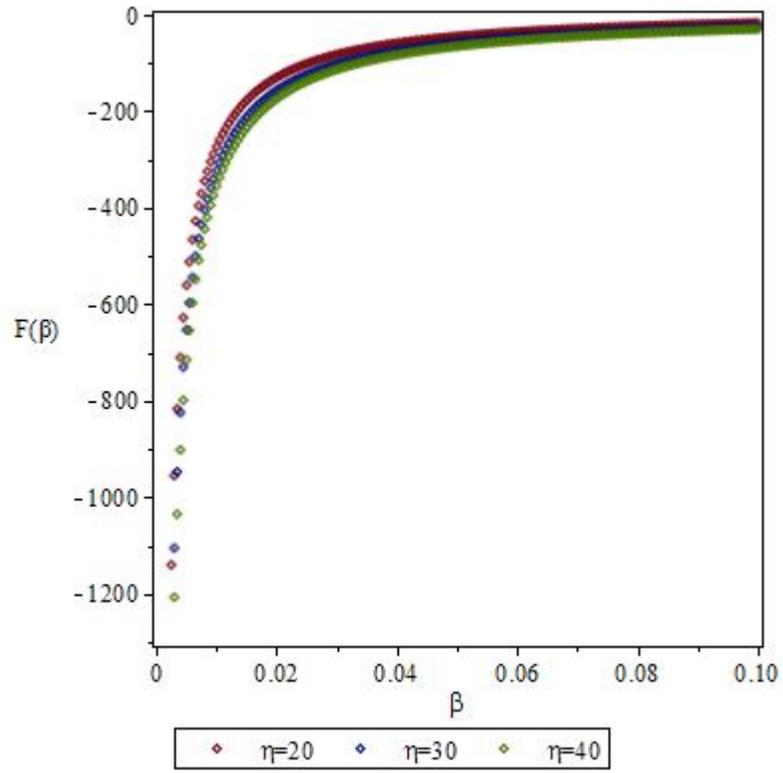

Fig.13: Vibrational free energy F as function of $\beta$ for different $\eta$ (20, 30 and 40)

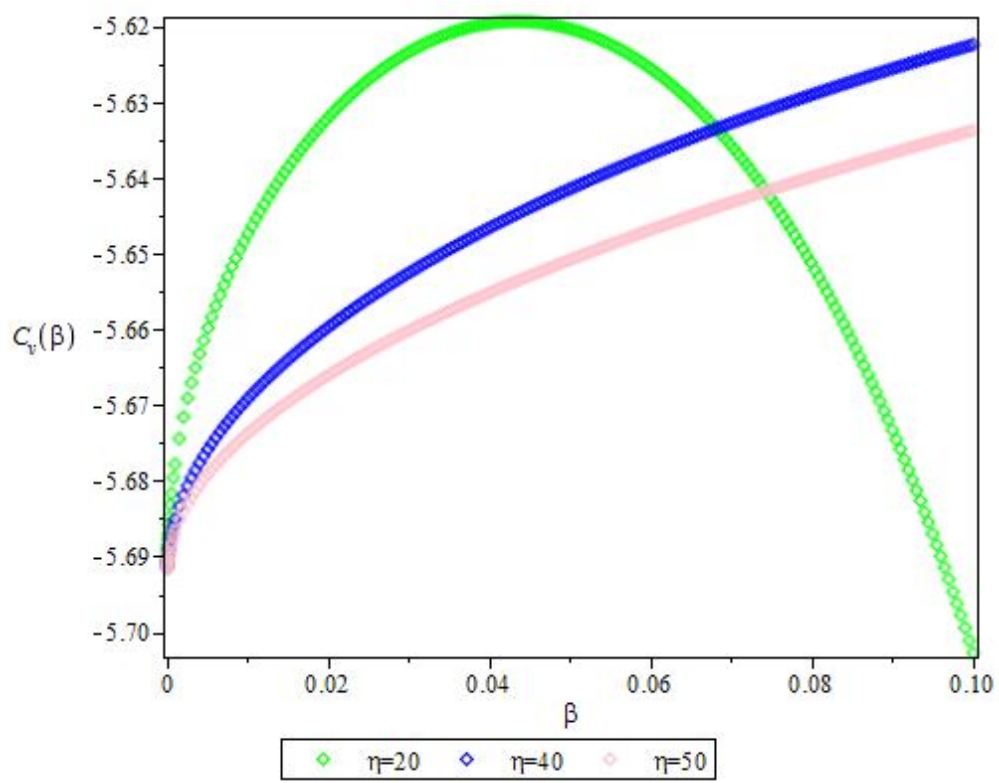

Fig.14: Vibrational specific capacity C as function of $\beta$ for different $\eta$ (20, 40 and 50)

Table 1: Relativistic Energy of the system for different states and dimensions and
$m = 2, p_1 = 1, p_2 = 2, p_3 = 1, \lambda = 4, \omega = 2, \alpha = 0.1$

| | $l$ | $E_{n,l}^D(D=1)$ | $E_{n,l}^D(D=2)$ | $E_{n,l}^D(D=3)$ | $E_{n,l}^D(D=4)$ | $E_{n,l}^D(D=5)$ | $E_{n,l}^D(D=6)$ |
|---|---|---|---|---|---|---|---|
| | 0 | 6.004372513, -2.005000032 | 5.776793059, -2.062560425 | 6.004372513, -2.005000032 | 6.620335419, -2.620334719 | 7.495604262, -3.495604077 | 17.12398757, -13.12398757 |
| | 1 | 6.004372513, -2.005000032 | 6.620335419, -2.620334719 | 7.495604262, -3.495604077 | 8.526676003, -4.526675934 | 9.650816933, -5.650816901 | 10.83256285, -6.832562831 |
| $n=0$ | 2 | 7.495604262, -3.495604077 | 8.526676003, -4.526675934 | 9.650816933, -5.650816901 | 10.83256285, -6.832562831 | 12.05161678, -8.051616768 | 13.29590633, -9.295906327 |
| | 3 | 9.650816933, -5.650816901 | 10.83256285, -6.832562831 | 12.05161678, -8.051616768 | 13.29590633, -9.295906327 | 14.55793242, -10.55793241 | 15.83284135, -11.83284135 |
| | 4 | 12.05161678, -8.051616768 | 13.29590633, -9.295906327 | 14.55793242, -10.55793241 | 15.83284135, -11.83284135 | 17.11737411, -13.11737411 | 18.40927075, -14.40927075 |
| | 0 | 6.014349165, -2.026014178 | 5.787369247, -2.081863559 | 6.014349165, -2.026014178 | 6.628984703, -2.628983997 | 7.502877993, -3.502877808 | 8.532801823, -4.532801753 |
| | 1 | 6.014349165, -2.026014178 | 6.628984703, -2.628983997 | 7.502877993, -3.502877808 | 8.532801823, -4.532801753 | 9.656043347, -5.656043316 | 10.83709038, -6.837090368 |
| $n=1$ | 2 | 7.502877993, -3.502877808 | 8.532801823, -4.532801753 | 9.656043347, -5.656043316 | 10.83709038, -6.837090368 | 12.05559545, -8.055595440 | 13.29944688, -9.299446878 |
| | 3 | 9.656043347, -5.656043316 | 10.83709038, -6.837090368 | 12.05559545, -8.055595440 | 13.29944688, -9.299446878 | 14.56111725, -10.56111725 | 15.83573272, -11.83573272 |
| | 4 | 12.05559545, -8.055595440 | 13.29944688, -9.299446878 | 14.56111725, -10.56111725 | 15.83573272, -11.83573272 | 17.12001984, -13.12001984 | 18.41170822, -14.41170822 |
| | 0 | 6.029267827, -2.055451751 | 5.803178383, -2.110487240 | 6.029267827, -2.055451751 | 6.641928412, -2.641927696 | 7.513770598, -3.513770412 | 8.541979796, -4.541979726 |
| | 1 | 6.029267827, -2.055451751 | 6.641928412, -2.641927696 | 7.513770598, -3.513770412 | 8.541979796, -4.541979726 | 9.663876287, -5.663876255 | 10.84387734, -6.843877327 |
| $n=2$ | 2 | 7.513770598, -3.513770412 | 8.541979796, -4.541979726 | 9.663876287, -5.663876255 | 10.84387734, -6.843877327 | 12.06156051, -8.061560498 | 13.30475563, -9.304755625 |
| | 3 | 9.663876287, -5.663876255 | 10.84387734, -6.843877327 | 12.06156051, -8.061560498 | 13.30475563, -9.304755625 | 14.56589299, -10.56589298 | 15.84006864, -11.84006864 |
| | 4 | 12.06156051, -8.061560498 | 13.30475563, -9.304755625 | 14.56589299, -10.56589298 | 15.84006864, -11.84006864 | 17.12398757, -13.12398757 | 18.41536374, -14.41536374 |